\documentclass[conference]{IEEEtran}
\IEEEoverridecommandlockouts
\usepackage{cite}
\usepackage{CJKutf8}
\usepackage{subfigure}
\usepackage{amsmath,amssymb,amsfonts}
\usepackage{algorithmic}
\usepackage{algorithm}
\usepackage{multirow}

\usepackage{graphicx}
\usepackage{textcomp}
\usepackage{xcolor}
\usepackage{xcolor}
\makeatletter

\newcommand{\Rmnum}[1]{\expandafter\@slowromancap\romannumeral #1@}
\makeatother
\usepackage{booktabs}
\def\BibTeX{{\rm B\kern-.05em{\sc i\kern-.025em b}\kern-.08em
T\kern-.1667em\lower.7ex\hbox{E}\kern-.125emX}}
\begin{document}

\title{
Steady-State Error Compensation for Reinforcement Learning with Quadratic Rewards
}

\author{\thanks{This work is supported by Guangzhou basic and applied basic research project under Grant 2023A04J1688.}\IEEEauthorblockN{Liyao Wang}
\IEEEauthorblockA{\textit{Shien-Ming Wu School of } \\
\textit{Intelligent Engineering} \\
\textit{South China University of Technology}\\
Guangzhou, China \\
202320160010@mail.scut.edu.cn}
\and
\IEEEauthorblockN{Zishun Zheng}
\IEEEauthorblockA{\textit{Shien-Ming Wu School of} \\
\textit{Intelligent Engineering} \\
\textit{South China University of Technology}\\
Guangzhou, China \\
202320160023@mail.scut.edu.cn}
\and
\IEEEauthorblockN{Yuan lin{*}}
\IEEEauthorblockA{\textit{Shien-Ming Wu School of } \\
\textit{Intelligent Engineering} \\
\textit{South China University of Technology}\\
Guangzhou, China \\
yuanlin@scut.edu.cn}
}
\maketitle
\begin{abstract}

The selection of a reward function in Reinforcement Learning (RL) has garnered significant attention because of its impact on system performance. Issues of significant steady-state errors often manifest when quadratic reward functions are employed. Although absolute-value-type reward functions alleviate this problem, they tend to induce substantial fluctuations in specific system states, leading to abrupt changes. In response to this challenge, this study proposes an approach that introduces an integral term. By integrating this integral term into quadratic-type reward functions, the RL algorithm is adeptly tuned, augmenting the system's consideration of reward history, and consequently alleviates concerns related to steady-state errors. Through experiments and performance evaluations on the Adaptive Cruise Control (ACC) and lane change models, we validate that the proposed method effectively diminishes steady-state errors and does not cause significant spikes in some system states.
\end{abstract}


\section{
Introduction
}

Reinforcement learning is a data-driven approach to decision-making\cite{arulkumaran2017brief}. In 2015, Mnih et al. pioneered the fusion of deep learning with reinforcement learning by introducing the Deep Reinforcement Learning (DRL) algorithm, which significantly augmented the representational capacity of reinforcement learning algorithms\cite{mnih2015human}. Today, this approach has extensive applications across various industries, including gaming, robotics, natural language processing, healthcare, Industry 4.0, smart grids, and beyond\cite{li2017deep}.

Optimal control theory is centered on crafting control policies to attain optimal performance for a given system based on a specific performance metric. This metric is commonly expressed as a cost or objective function, and the objective of optimal control is to discover a control policy that minimizes or maximizes this performance metric\cite{lewis2012optimal}.

In control engineering, RL is adept at uncovering optimal control laws. Through iterative interactions with the environment, optimal control strategies can be identified, relying instead on a trial-and-error approach. Reinforcement Learning is conceptualized as a Markov Decision Process (MDP), where the agent (controller) observes the state of the environment, takes actions, receives rewards, and refines its policy based on feedback. Through learning and optimization, reinforcement learning progressively unveils the optimal control law, empowering the system to attain optimal performance in specific tasks or goals\cite{sutton1999reinforcement}. The reward function in reinforcement learning is akin to the cost function in control theory, defining the agent's objectives in the environment and exerting direct influence over the algorithm's performance and convergence speed\cite{matignon2006reward}.

Engel et al. established a connection between the reward function and the ultimate control performance of closed-loop systems. Their research revealed that a reward function grounded in quadratic values could result in notable steady-state errors in the final state variables. Furthermore, their exploration of absolute value-based reward functions demonstrated a decrease in steady-state errors \cite{engel2014line}.

Weber et al. proposed an integral action state augmentation for arbitrary actor-critic based-RL controllers, which can reduce the steady-state error in reference and disturbance rejection tracking control problems. The approach doubles the number of output neurons within the actor (one additional integral term per
output), and each output neuron pair will route one neuron to the
environment and the other to the integrator unit. Furthermore, it uses a steady-state error compensation specific reward to enhance the learning process.\cite{weber2023steady} Different from the methods mentioned above, this paper focuses on integrating the error integral into the reward function without changing the structure of the neural network, which is easier to implement.

The primary contributions of this study can be summarized as follows: This study identified that the reward function based on the absolute value still induces considerable fluctuations in specific state changes throughout the entire control process, resulting in jerky control performance. Nevertheless, the introduction of an integral term helps diminish the steady-state error of the quadratic function without inflicting significant spikes of certain system states.

The subsequent sections of this paper are as follows: Section \Rmnum{2} introduces the basic theory of reinforcement learning and the Deep Deterministic Policy Gradient (DDPG) algorithm. Section \Rmnum{3} outlines the specific method for introducing the integral term. Section \Rmnum{4} details the modeling of the experimental environment and conducts simulations to validate the proposed method. Section \Rmnum{5} summarizes the entire paper.

\section{
PRELIMINARY
}

\subsection{
Reinforcement Learning
}

An MDP serves as the ideal mathematical model for RL\cite{franccois2018introduction}. In a standard RL process, it is initially assumed that the environment $e$ interacts with the agent. At each time step $t$, the agent receives observations $s$ from the environment, takes action $a$, causing the environment to change on the basis of the agent's actions, and receives an immediate reward $r$. In this context, we consider reinforcement learning to be entirely observable, allowing the agent to access global information.

The agent's actions are determined by policy $\pi$. The policy $\pi(a|s)=P(A_t=a|S_t=s)$ is a function that expresses the probability of taking action $a$ given the input state $s$. When a policy is deterministic, it outputs only one specific action with a probability of one at each state, whereas the probabilities of other actions are 0. In the case of a stochastic policy, it outputs a probability distribution over actions at each state, and  action can be sampled on the basis of this distribution. Probability of the agent transitioning from state $s$ to the next state
$s'$ after taking action $a$ according to policy $\pi$ is defined as $P(s'|s)=P(s_{t+1}=s'|s_t=s)$, representing the state transition probability with Markovian properties. In an MDP, starting from time step 0 until reaching the termination state, the sum of discounted rewards is referred to as the return.
\begin{equation}
\begin{aligned}
G_t=R_t+\gamma R_{t+1}+\gamma^2R_{t+2}+\cdots=\sum_{k=0}^{\infty}\gamma^kR_{t+k}
\end{aligned}
\end{equation}

Here, $\gamma$ is the discount factor, reflecting the ratio between the future reward value and the current reward value.

Furthermore, the definition of maximizing the cumulative expected return $E[G_t|S_t=s]$ is introduced. Considering the presence of actions in MDP, the action-value function $Q^{\pi}(s,a)$ is defined:
\begin{equation}
\begin{aligned}
Q^{\pi}(s,a)=E_{\pi}[G_t|S_t=s,A_t=a]
\end{aligned}
\end{equation}

It represents the cumulative expected return when performing action $a$ in a certain state $s$. The Bellman equation is then derived as
\begin{equation}
\begin{aligned}
Q^{\pi}(s,a)&=E_{\pi}[R_t+\gamma Q^{\pi}(S_{t+1},A_{t+1})|S_t=s,A_t=a]\\
&=r(s,a)+\gamma \sum_{s'\in S}p(s'|s,a)\sum_{a'\in A}\pi(a'|s')Q^{\pi}(s',a')
\end{aligned}
\end{equation}

It can be used for iterative solutions to obtain the action-value function $Q^{\pi}(s,a)$. Ultimately, the core objective of reinforcement learning is to determine an optimal policy
\begin{equation}
\begin{aligned}
\pi^*=\arg \max E_{\pi}[\sum_{t=0}^{\infty}\gamma^tr_t(s_t,a_t)]
\end{aligned}
\end{equation}

It can maximize the action-value function $Q^{\pi}(s,a)$.

\subsection{
DDPG Algorithm
}

The DDPG algorithm combines concepts from deterministic policy gradient algorithms\cite{lilicrap2016continuous}. Built upon the actor-critic framework, it employs deep neural networks to approximate both the policy network and the action-value function. Training of the policy network and value network model parameters is accomplished through batch gradient descent. The algorithm adopts a dual neural network architecture for both the policy and value functions, consisting of online and target networks, enhancing stability during the learning process and expediting convergence. In addition, the algorithm incorporates an experience replay mechanism, where data generated by the actor interacting with the environment is stored in an experience pool. Batch data samples are then extracted for training, similar to the experience replay mechanism in Deep Q-Network, with the aim of eliminating sample correlations and dependencies to improve convergence.

\section{
method
}
\subsection{
Problem Statement
}
The quadratic reward function of RL is defined as follows:
\begin{equation}
\begin{aligned}
C_{qua} = -\left[(\mathbf{x}_{t+1}-\mathbf{x}_g)^T\mathbf{L}(\mathbf{x}_{t+1}-\mathbf{x}_g) \right. \\
\left. + \mathbf{u}_t^{\top}\mathbf{M}\mathbf{u}_t + \dot{\mathbf{x}}_{t+1}^{\top}\mathbf{N}\dot{\mathbf{x}}_{t+1}\right]
\end{aligned}
\end{equation}

In this context, $x_{t+1}$ and $u_{t} $ is the state and action, $\mathbf{x_g}$ represents the desired state value and is a constant conference. $\mathbf{L}, \mathbf{M}, \mathbf{N}$ are weight matrices. These matrices are used to fine-tune the relative impacts of the state, action vector, and rate of state changes on the overall cost. Among these, the rate of state changes considers factors such as comfort in specific control problems. For instance, in the context of a moving car, if the acceleration change rate $jerk$ is too high, it will make passenger uncomfortable. 
In contrast to the cost function in optimal control, where the objective is to minimize a cost function, reinforcement learning strives to maximize expected returns.

Similarly, the absolute value reward function is defined as:

\begin{equation}
\begin{aligned}
C_{abs} = -\left(\mathbf{O}|(\mathbf{x}_{t+1}-\mathbf{x}_g)| + \mathbf{P}|\mathbf{u}_t| + \mathbf{Q}|\dot{\mathbf{x}}_{t+1}|\right)
\end{aligned}
\end{equation}

Where $\mathbf{O},\mathbf{P},\mathbf{Q}$ are the weight matrices.
After determining the specific reward functions, experiments were conducted using both the quadratic and absolute value reward functions, as illustrated in Section \Rmnum{4} Simulation. The results clearly demonstrate that, for the quadratic reward function, although it achieves relatively smooth control, the difference between $\mathbf{x}_{t+1}$ and $\mathbf{x}_g$ is larger than that for the absolute value reward function at steady-state. On the other hand, the absolute value reward function, while achieving smaller steady-state errors, exhibits higher fluctuation in the state during the control process.

To address this issue, we propose two new reward functions based on integral terms:
\subsection{
Solution
}
The two reward functions presented in this study are both derived from the quadratic reward function, chosen for its superior smoothness, which helps avoid significant peaks in the system's variables. The fundamental concept is to mitigate the steady-state error associated with the quadratic reward function. Analogous to the incorporation of integral terms in PID control to diminish steady-state errors in control systems, this study aggregates the system's steady-state errors and integrates them as a novel penalty into the reinforcement learning reward function.

Therefore, the reward function incorporating the steady-state error is defined as:

\begin{equation}
\begin{aligned}
C_{pi}=C_{qua}+C_{I}\\
C_{I}=(\mathbf{c}_{I})^{\top}\mathbf{Z}\mathbf{c}_{I}
\end{aligned}
\end{equation}

$\mathbf{c}_{I}$ represents the accumulated error term. $Z$ is the weight value. Through experiments, it has been observed that if the accumulated errors throughout the entire control process are directly summed up, the training process becomes very slow and sometimes struggles to converge. Therefore, two methods are proposed to calculate the accumulated error term. 
\subsubsection{Method 1}

The first method is as follows:

\begin{equation}
\mathbf{c}_I=\left\{
\begin{array}{lr}
\enspace\quad\quad\quad\quad 0 \quad\quad\quad \quad    \enspace   if\quad0<t< t_{threshold}\\
\sum_{t_{threshold}}^{t}\left\vert \mathbf{x}_{t}-\mathbf{x}_g\right\vert       \quad if \quad t\geq t_{threshold}
\end{array}
\right.
\end{equation}

Where $\left\vert \mathbf{x}_{t}-\mathbf{x}_g\right\vert$ is used to describe the error value at time $t$. The variable $t_{threshold}$ is defined as a threshold, and its setting depends on when the system enters a steady-state. This approach only sums up the smaller steady-state errors after stabilization. Therefore, it allows the use of relatively large weight values $Z$ to reduce steady-state errors. Meanwhile, it significantly improves the convergence speed of reinforcement learning.

\subsubsection{Method 2}

In most cases, if the approximate time to reach steady-state is unknown in advance and cannot be obtained through experimentation, the first method may fail to determine its critical value. Hence, we introduce the second method:
\begin{equation}
\begin{aligned}
\mathbf{c}_I =\sum_{0}^{t}\kappa(t) \left\vert \mathbf{x}_{t}-\mathbf{x}_g\right\vert
\end{aligned}
\end{equation}

Here, $\kappa(t)$ is a weight factor that can be artificially set depending on the environment. Its purpose is to ensure that the RL strategy is not excessively influenced by accumulated transient change errors, thus accelerating convergence. Therefore,  $\kappa(t)$ should gradually increase over time steps, aligning with our increasing emphasis on steady-state errors. In this study, we use the $\mathtt{sigmoid}$ function to represent the $\kappa(t)$ term\cite{elfwing2018sigmoid}:
\begin{equation}
\begin{aligned}
\kappa(t)=\frac{1}{1+ae^{T/2-t}}
\end{aligned}
\end{equation}

$T$ is the length of an episode. The scaling of the $\mathtt{sigmoid}$ function can be controlled by adjusting its parameters, typically denoted as $a$.

\section{
Simulation
}

This section primarily introduces the two models applied to validate the proposed two solutions: the ACC and lane change models. In addition, the presentation of the experimental results is discussed.
\subsection{
ACC
}
\subsubsection{Model}

\begin{figure}[htbp]
\centerline{\includegraphics[width=0.5\textwidth]{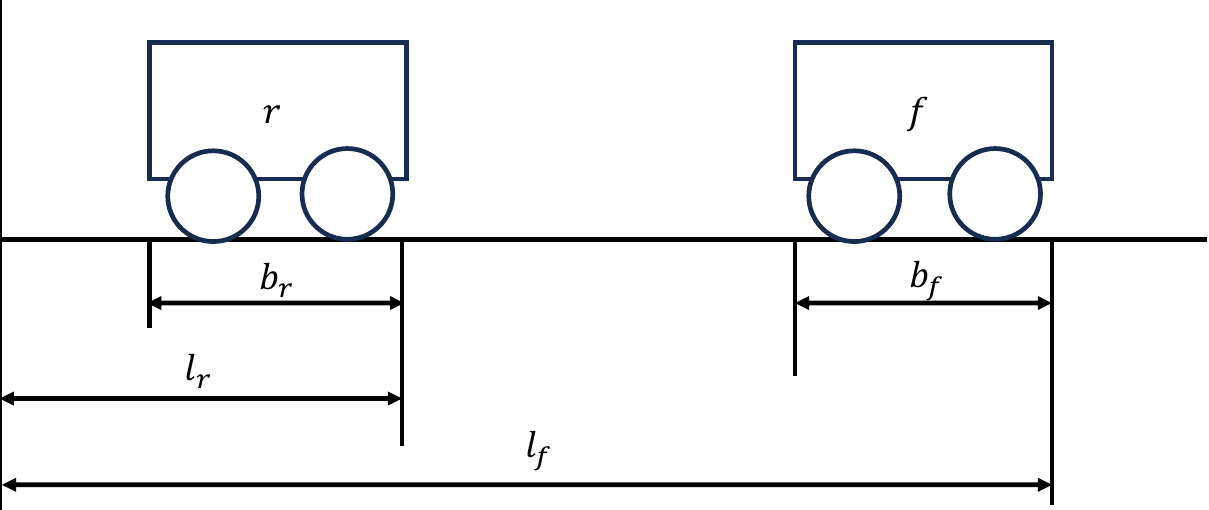}}
\caption{Schematic for two-car following.}
\label{fig1}
\end{figure}

The schematic of the ACC is shown in Fig. \ref{fig1}.

Where $f$ represents the front vehicle, $r$ is the controlled vehicle, $b_f$ is the length of the target vehicle, $b_r$ is the length of the controlled vehicle, $l_r$ is the distance traveled by the controlled vehicle, and $l_f$ is the distance traveled by the target vehicle. The equations for the spacing and velocity errors between the target and controlled vehicles are as follows:

\begin{equation}
\begin{aligned}
&\ d_{safe}=d_0+hv_r\\
&\ d_{rf}=d_{f}-d_r-l_{f}\\
&\ e=d_{rf}-d_{safe} \\
&\ e_v=v_{f}-v_r,
\end{aligned}
\end{equation}

Where $d_0$ is the standstill safety distance and $h$ is the headway time. This study adopts a first-order inertia system to approximate the vehicle motion process, leading to the following state-space equations\cite{lin2020comparison}:

\begin{equation}
\begin{aligned}
&\ \dot e=e_v-ha_r \\
&\ \dot e_v=a_{f}-a_r=-a_r \\
&\ \dot a_r=\frac{u_r-a_r}{\tau}
\end{aligned}
\end{equation}

Where $a_r$ is the actual acceleration of the controlled vehicle, $a_f$ is the actual acceleration of the target vehicle, $u_f$ is the system output representing the desired acceleration of the controlled vehicle, and $\tau$ is the constant of the first-order inertia system. To minimize variables that could cause steady-state errors, this study sets $a_f=0$, indicating that the preceding vehicle is moving at a constant speed.

For the RL in the ACC environment, the state variables are set as ($e, e_v, a_r$), the output actions as $u_r$, and the initial state is configured as (5, 5, 0).

\subsubsection{
Reward function
}

Here, only the terms corresponding to nonzero weights in the previous reward function are mentioned. Therefore, the four reward functions are defined as follows:

\begin{equation}
\begin{aligned}
C_{qua} = w_1(\frac{e_t}{e_{nmax}})^2+w_2(\frac{u_r}{u_{max}})^2\\
+w_3(\frac{\dot a_r}{(u_{max}-u_{min})/\tau})^2
\end{aligned}
\end{equation}
\begin{equation}
\begin{aligned}
C_{abs}= w_1\left\vert \frac{e_t}{e_{nmax}}
\right\vert+w_2\left\vert \frac{u_r}{u_{max}} \right\vert \\
+w_3\left\vert \frac{\dot a_r}{(u_{max}-u_{min})/\tau} \right\vert
\end{aligned}
\end{equation}

The PI-reward function is defined as $
C_{pi}=C_{qua}+C_{I}$,
$C_{I}=\omega_4(\frac{c_{I}}{c_{nmax}})^2$. $C_{I1}$ and $C_{I2}$ correspond to Methods 1 and 2, respectively:
\begin{equation}
\mathbf{c}_{I1}=\left\{
\begin{array}{lr}
\enspace\quad\quad\quad\quad 0 \quad\quad\quad \quad       if\quad0<t< t_{threshold}\\
\sum_{t_{threshold}}^{t}\left\vert e_t\right\vert     \enspace \quad   \quad if \quad t\geq t_{threshold}
\end{array}
\right.
\end{equation}
\begin{equation}
\begin{aligned}
\mathbf{c}_{I2} =\sum_{0}^{t}\kappa(t) \left\vert e_t\right\vert
\end{aligned}
\end{equation}

The first term $e_{t}$ corresponds to the spacing error in the ACC system. The second term represents the penalty for actions, while the third term corresponds to the acceleration. $\omega_1,\omega_2,\omega_3,\omega_4$ denotes the weights, and the remaining coefficients $u_{max}$, $u_{min}$, $e_{nmax}$ and $c_{nmax}$ are normalization factors. The specific values for all coefficients are provided in the Table \ref{tab1}.

\begin{table}[!t]
\caption{
ACC parameter values
}
\centering
\begin{tabular}{cccc}
\toprule
\textbf{Parameter} & \textbf{Values}\\
\midrule
Constant time gap $h$ & 1s\\
Time step $t$ & 0.1s\\
Allowed maximum control input $u_{max}$ & 2m/$s^2$\\
Allowed minimum control input $u_{min}$ & -3m/$s^2$\\
Nominal maximum gap-keeping error $e_{nmax}$ & 15m\\
$c_{nmax}$ & 60m\\
$t_{threshold}$ & 125\\
$\tau$ & 0.4\\
$a$ in $\kappa(t)$ & $\frac{1}{10}$\\
$\omega_1=\omega_2=\omega_3$ & $\frac{1}{3}$\\
$\omega_4$ for $PI1$ & 0.1\\
$\omega_4$ for $PI2$ & 0.05\\
\bottomrule
\end{tabular}
\label{tab1}
\end{table}

\subsubsection{
Result
}

\begin{figure}[htbp]
\centerline{\includegraphics[width=0.5\textwidth]{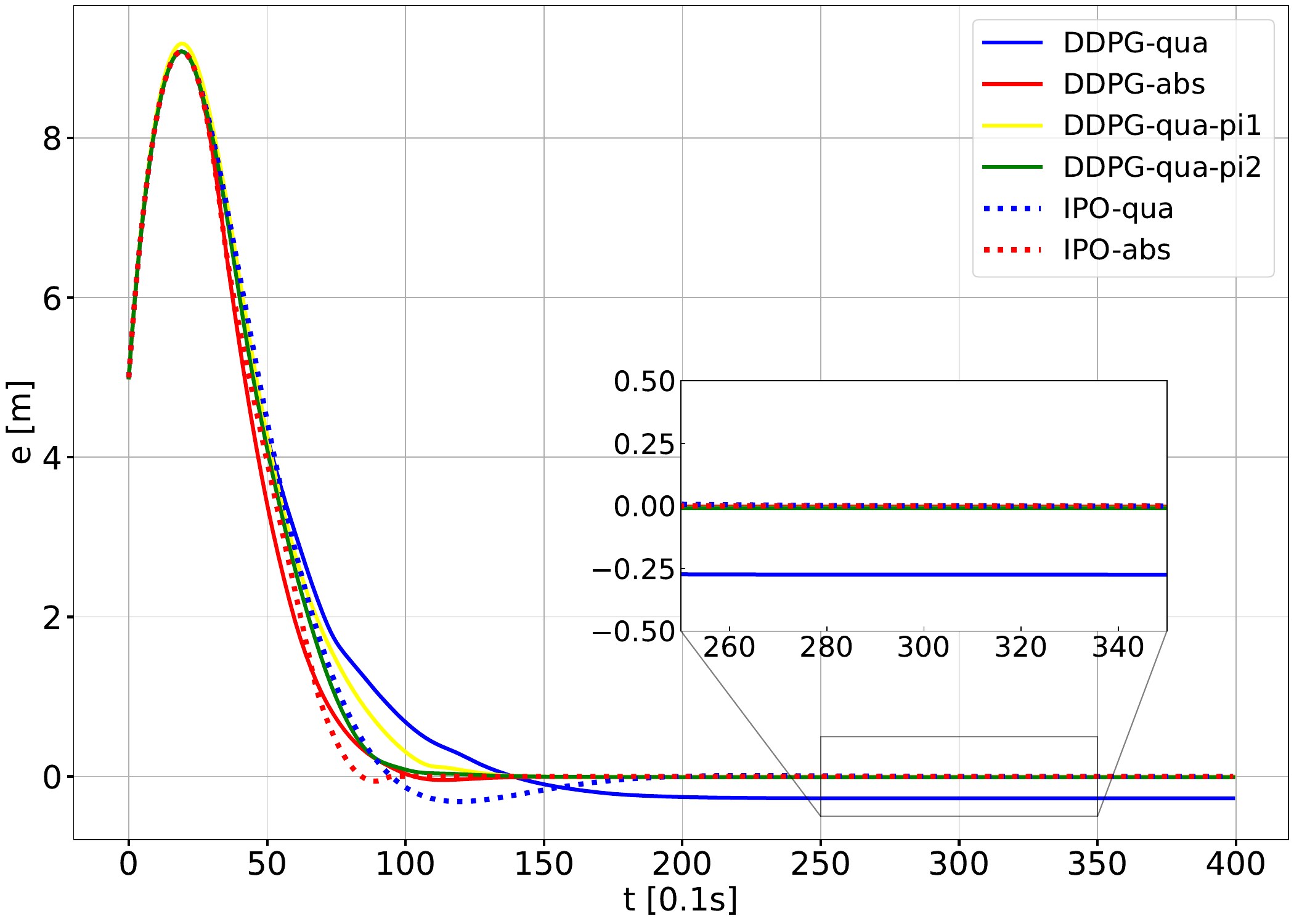}}
\caption{Spacing errors in the ACC model. (Only the first 400 timesteps are shown, and the actual training is 600 timesteps) }
\label{fig3}
\end{figure}

\begin{figure}[htbp]
\centerline{\includegraphics[width=0.5\textwidth]{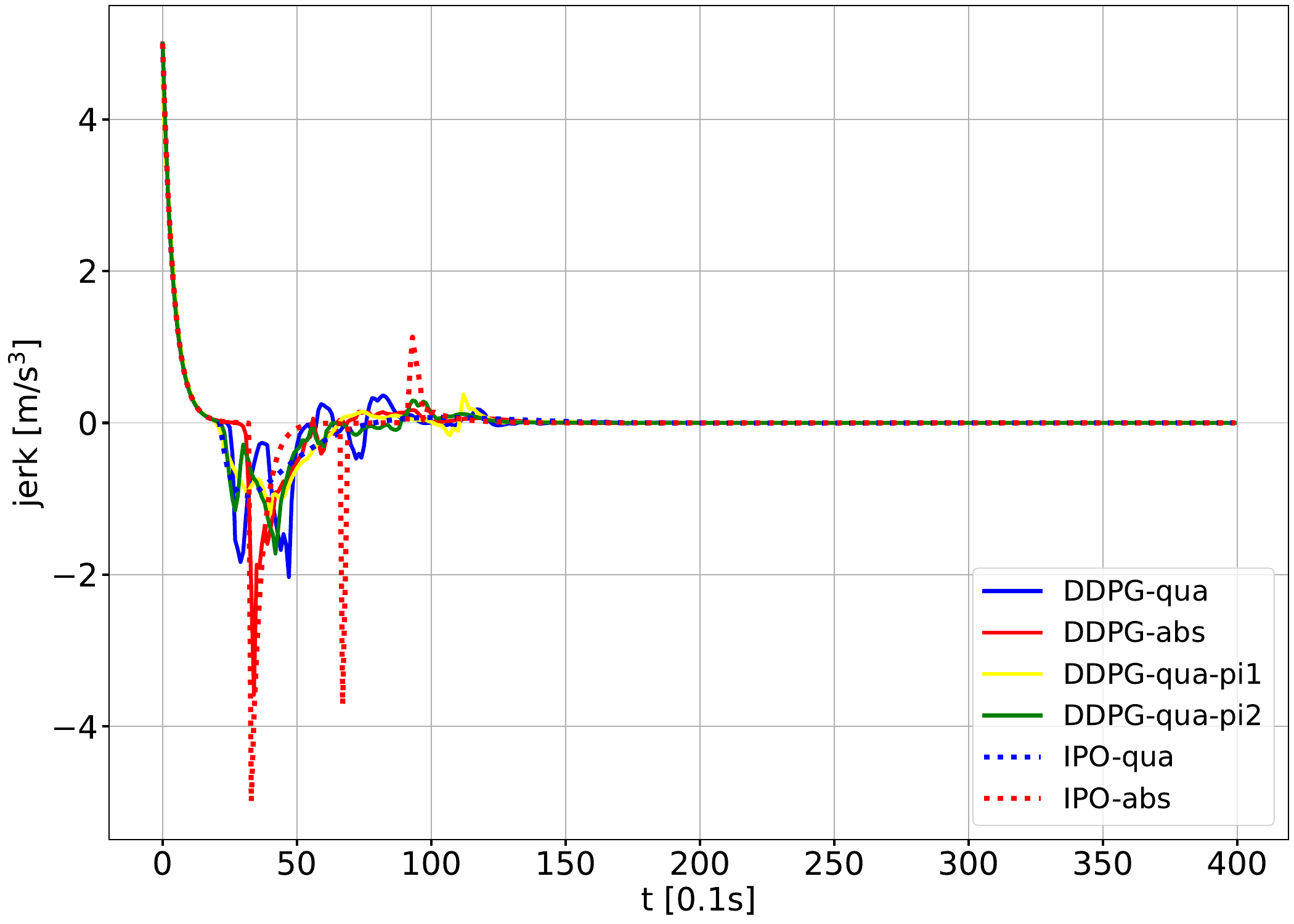}}
\caption{The rate of change in vehicle acceleration for ACC. (Only the first 400 timesteps are shown, and the actual training is 600 timesteps)}
\label{fig4}
\end{figure}

First, the results of the ACC model are depicted in Fig. \ref{fig3} and Fig. \ref{fig4}, where the IPO method illustrates the optimal curve, whereas reinforcement learning often falls short of achieving such performance. The primary focus is on the final spacing error and acceleration change rate during the tracking process. Using the quadratic reward function leads to a larger steady-state spacing error, as shown in Fig. \ref{fig3}, and the absolute value reward function induces significant peaks in the acceleration change rate throughout the control process, as shown in Fig. \ref{fig4}. However, employing PI-based Methods 1 and 2 can reduce the final steady-state spacing error and result in small fluctuations in the acceleration change rate. The specific steady-error values and undiscounted episodic rewards (costs) are shown in Table \ref{tab3}.

\begin{table}[!t]
\caption{
ACC date
}
\centering
\begin{tabular}{cccc}
\toprule
\multirow{2}{*}{\textbf{Method}} & \multirow{2}{*}{\textbf{Steady-state errors [m]}}& \textbf{Undiscounted episodic}\\
&&\textbf{rewards (costs)}\\
\midrule
IPO-qua& 4.5e-7& -9.12\\
DDPG-qua& 2.7e-1& -9.24\\
DDPG-qua-pi1& 7.3e-3& -9.30\\
DDPG-qua-pi2& 9.2e-3& -9.44\\
IPO-abs& 1.1e-9& -19.79\\
DDPG-abs& 2.9e-3& -20.30\\
\bottomrule
\end{tabular}
\label{tab3}
\end{table}


\subsection{Lane\enspace change\enspace control}
\subsubsection{Model}

\begin{figure}[htbp]
\centerline{\includegraphics[width=0.4\textwidth]{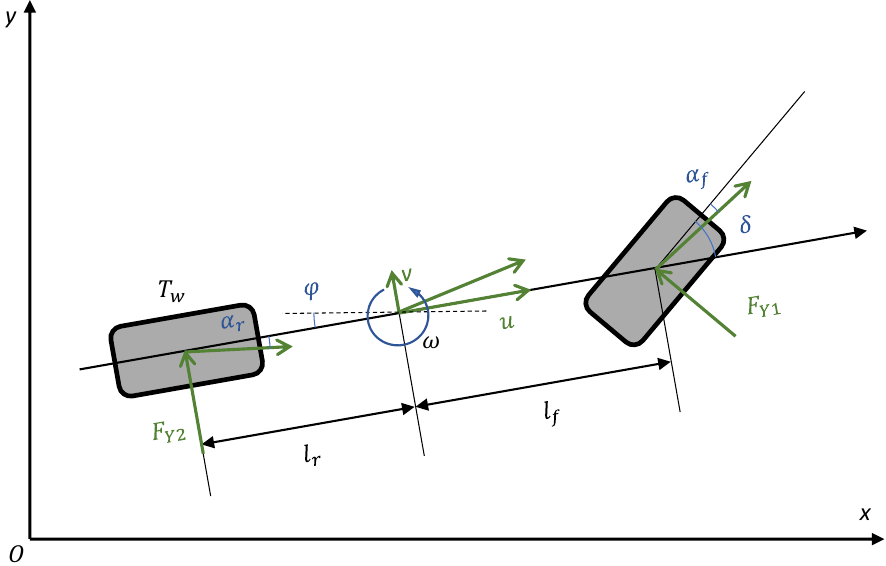}}
\caption{Dynamic bicycle model for lane change\cite{ge2021numerically}.}
\label{fig2}
\end{figure}

The vehicle dynamics model used for the lane change in this study is illustrated in Fig. \ref{fig2}.

The state-space equations are as follows:

\begin{equation}
\begin{aligned}
&\ \dot x=u\cos{\varphi}-v\sin{\varphi} \\
&\ \dot y=v\cos{\varphi}+u\sin{\varphi} \\
&\ \dot \varphi=\omega \\
&\ \dot u=v\omega-\frac{1}{m}F_{Y1}\sin{\delta} \\
&\ \dot v=-u\omega+\frac{1}{m}(F_{Y1}\cos{\delta}+F_{Y2}) \\
&\ \dot \omega=\frac{1}{I_z}(l_fF_{Y1}\cos{\delta}-l_rF_{Y2}) \\
&\ \dot \delta_{old}=\frac{\delta - \delta_{old}}{\tau}
\end{aligned}
\end{equation}

Where $x$ and $y$ represent the distances of the vehicle along the $x$ and $y$ axes, $\varphi$ is the angle between the vehicle body and the x-axis, $u$ is the velocity of the vehicle center of mass along the body, $v$ is the velocity of the vehicle center of mass perpendicular to the body, $\omega$ is the angular velocity of the vehicle body, and $\delta_{old}$ stores the decision $\delta$ made by the vehicle in the previous time step. $F_{Y1}$ and $F_{Y2}$ are the lateral forces. Under a mild steering angle, we can assume that:
\begin{equation}
\begin{aligned}
&\ F_{Y1} \approx k_f(\frac{v+l_f\omega}{u})-\delta\\
&\ F_{Y2} \approx k_r\frac{v-l_r\omega}{u}
\end{aligned}
\end{equation}

The lane change scenario considered in this paper involves the simplest case. The controlled vehicle smoothly approaches the centerline of the second lane. Therefore, the vehicle's longitudinal acceleration is defined as 0, and the state variable $x$ for longitudinal motion is not considered during the control process.

The distance of the vehicle in the y-direction from the centerline of the second lane is defined as $e$, which serves as a measure of the steady-state error in vehicle control.

For the RL in the lane change environment, the state variables are set as ($e, \varphi,u,v,\omega,\delta_{old}$), and the output actions as $\delta$. The initial state is configured as (4, 0, 30, 0, 0, 0).

\subsubsection{
Reward function
}

Similar to the ACC model, only the terms corresponding to nonzero weights in the previous reward function are mentioned. Therefore, the four reward functions are defined as follows:

\begin{equation}
\begin{aligned}
C_{qua} = w_1(\frac{y_t}{y_{nmax}})^2+w_2(\frac{\delta}{\delta_{max}})^2\\
+w_3(\frac{\dot \delta_{old}}{(\delta_{max}-\delta_{min})/\tau})^2
\end{aligned}
\end{equation}
\begin{equation}
\begin{aligned}
C_{abs}= w_1\left\vert \frac{y_t}{y_{nmax}}
\right\vert+w_2\left\vert \frac{\delta}{\delta_{max}} \right\vert \\
+w_3\left\vert \frac{\dot \delta_{old}}{(\delta_{max}-\delta_{min})/\tau} \right\vert
\end{aligned}
\end{equation}

The PI-reward function is defined as $
C_{pi}=C_{qua}+C_{I}$,
$C_{I}=\omega_4(\frac{c_{I}}{c_{nmax}})^2$. $C_{I1}$ and $C_{I2}$ correspond to Methods 1 and 2, respectively:
\begin{equation}
\mathbf{c}_I=\left\{
\begin{array}{lr}
\enspace\quad\quad 0 \quad\quad\quad \quad\quad\quad       if\quad0<t< t_{threshold}\\
\sum_{t_{threshold}}^{t}\left\vert y_t\right\vert       \quad \quad \quad if \quad t\geq t_{threshold}
\end{array}
\right.
\end{equation}
\begin{equation}
\begin{aligned}
\mathbf{c}_I =\sum_{0}^{t}\kappa(t) \left\vert y_t\right\vert
\end{aligned}
\end{equation}

The first term $y_{t}$ corresponds to the distance of the vehicle from the centerline of the second lane in the lane-changing system. The second term represents the penalty for actions, and the third term corresponds to the rate of angle change in the lane-changing system. $\omega_1,\omega_2,\omega_3,\omega_4$ denotes the weights, and the remaining coefficients 
 $\delta_{max}$, $\delta_{min}$, $y_{nmax}$ and $c_{nmax}$ are normalization factors. The specific values for all coefficients are provided in Table \ref{tab2}, and the values of $m,I_z,k_f,k_r,l_f$ and $l_r$ are referenced in \cite{zheng2024highway}.

\begin{table}[!t]
\caption{
Lane change parameter values
}
\centering
\begin{tabular}{cccc}
\toprule
\textbf{Parameter} & \textbf{Values}\\
\midrule
Mass of the vehicle \(m\) & 1470kg \\
Time step \(t\) & 0.1s \\
Yaw inertia of vehicle body \(I_z\) & 2400kg*$m^2$ \\
Front axle equivalent sideslip Stiffness \(k_f\) & -100000N/rad \\
Rear axle equivalent sideslip Stiffness \(k_r\) & -100000N/rad \\
Centroid to front axle distance \(l_f\) & 1.085m \\
Centroid to rear axle distance \(l_r\) & 2.503m \\
Allowed maximum control input \(\delta_{max}\) & \(5^\circ\) \\
Allowed minimum control input \(\delta_{min}\) & \(-5^\circ\) \\
Nominal maximum lateral error \(y_{nmax}\) & 4m \\
Distance between two centerlines $d$ & 4m \\
$c_{nmax}$ & 15m\\
$\tau$ &0.1 \\
\(t_{threshold}\) & 30 \\
\(a\) in \(\kappa(t)\) & \(\frac{1}{10}\) \\

\(\omega_1 = \omega_2 = \omega_3\) & \(\frac{1}{3}\) \\
\(\omega_4\) for PI1 & 0.1 \\
\(\omega_4\) for PI2 & 0.5 \\
\bottomrule
\end{tabular}
\label{tab2}
\end{table}

\subsubsection{
Result
}

\begin{figure}[htbp]
\centerline{\includegraphics[width=0.5\textwidth]{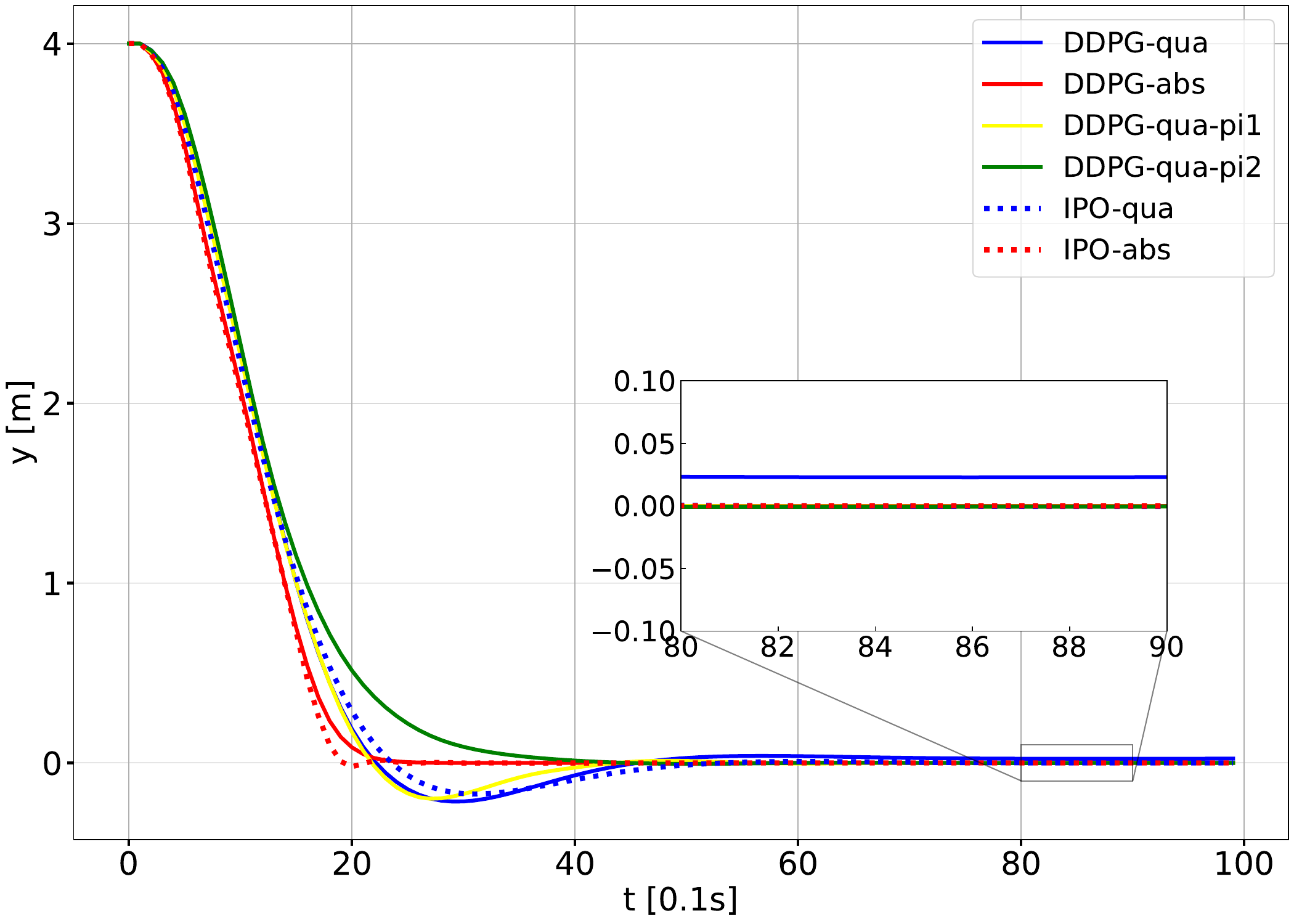}}
\caption{The distance between the car and the centerline in lane change. (Only the first 100 timesteps are shown, and the actual training is 150 timesteps)}
\label{fig5}
\end{figure}

\begin{figure}[htbp]
\centerline{\includegraphics[width=0.5\textwidth]{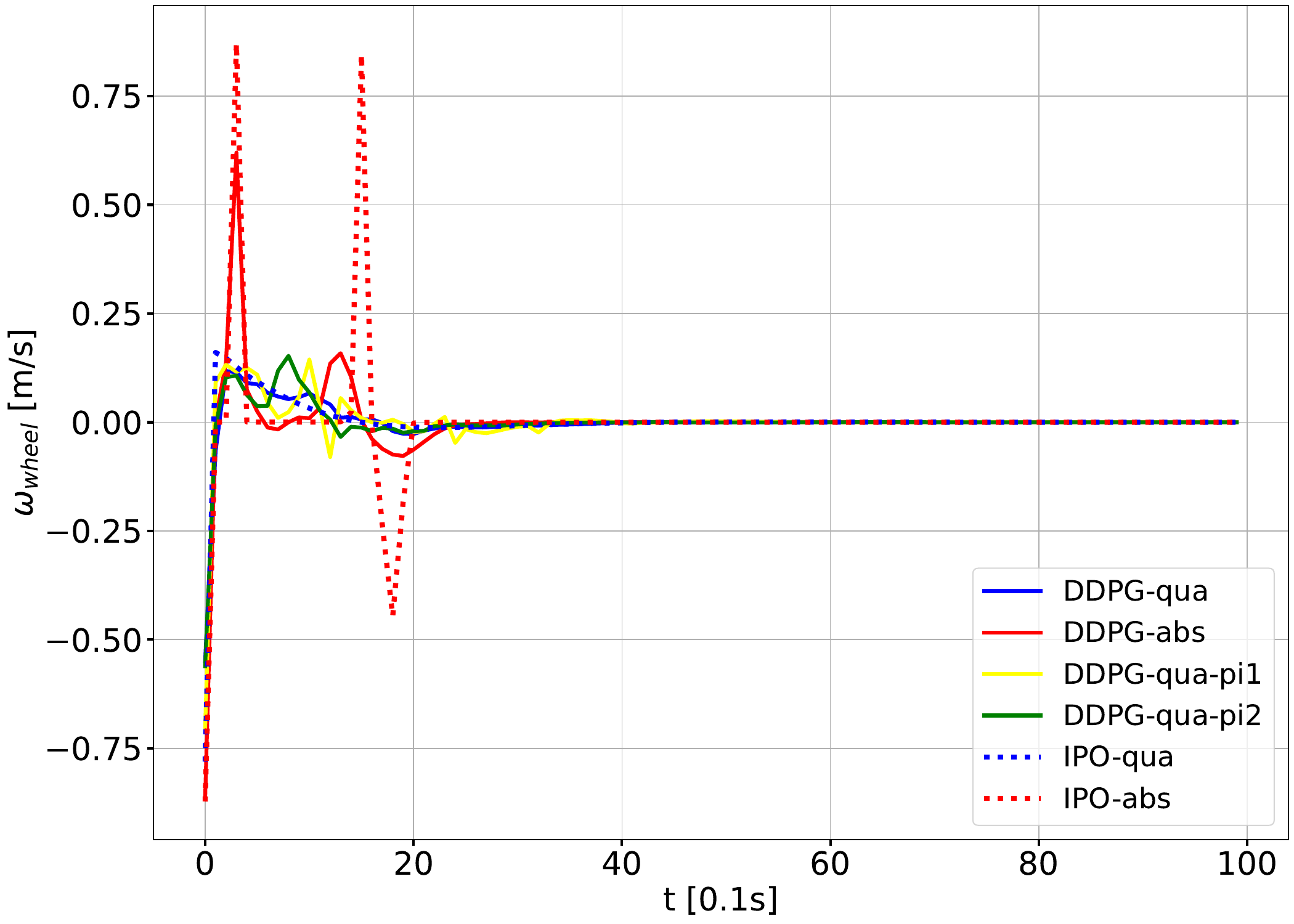}}
\caption{Rate of change in wheel angle for lane change. (Only the first 100 timesteps are shown, and the actual training is 150 timesteps) }
\label{fig6}
\end{figure}

For the lane change model results, as illustrated in Fig. \ref {fig5} and Fig. \ref{fig6}, the focus is on the final distance from the vehicle to the centerline of the second lane and the fluctuation in the wheel angle throughout the entire control process. Comparing the quadratic reward function with the absolute value reward function, as shown in Fig. \ref{fig5}, there is a difference of three orders of magnitude in the final distance steady-state error, as indicated in Table \ref{tab4}. In addition, the absolute value reward function exhibits larger fluctuations in the wheel angle change rate, as depicted in Fig. \ref{fig6}. The use of PI-based Methods 1 and 2 with quadratic rewards demonstrate superior control performance. The specific steady-state error values and undiscounted episodic rewards (costs) are reflected in Table \ref{tab4}.


\begin{table}[!t]
\caption{
Lane change date
}
\centering
\begin{tabular}{cccc}
\toprule
\multirow{2}{*}{\textbf{Method}} & \multirow{2}{*}{\textbf{Steady-state errors [m]}}& \textbf{Undiscounted episodic}\\
&&\textbf{rewards (costs)}\\
\midrule
IPO-qua& 2.4e-6& -3.80\\
DDPG-qua& 2.3e-2& -3.81\\
DDPG-qua-pi1& 1.8e-4& -3.86\\
DDPG-qua-pi2& 4.9e-4& -3.92\\
IPO-ab& 4.5e-7& -6.24\\
DDPG-abs& 5.5e-5& -6.26\\
\bottomrule
\end{tabular}
\label{tab4}
\end{table}

\section{
conclusion
}

This study utilizes PI-based quadratic reward functions by adding an integral term to reduce the steady-state errors of reinforcement learning. The results indicate that using a quadratic reward function may lead to significant steady-state errors in certain systems, whereas an absolute value reward function may result in an unsmooth control process with spikes. Adopting PI-based quadratic reward functions reduces the steady-state errors without inflicting significant spikes of certain system states.
\bibliographystyle{IEEEtran}
\bibliography{conference}

\begin{thebibliography}{10}
\providecommand{\url}[1]{#1}
\csname url@samestyle\endcsname
\providecommand{\newblock}{\relax}
\providecommand{\bibinfo}[2]{#2}
\providecommand{\BIBentrySTDinterwordspacing}{\spaceskip=0pt\relax}
\providecommand{\BIBentryALTinterwordstretchfactor}{4}
\providecommand{\BIBentryALTinterwordspacing}{\spaceskip=\fontdimen2\font plus
\BIBentryALTinterwordstretchfactor\fontdimen3\font minus \fontdimen4\font\relax}
\providecommand{\BIBforeignlanguage}[2]{{%
\expandafter\ifx\csname l@#1\endcsname\relax
\typeout{** WARNING: IEEEtran.bst: No hyphenation pattern has been}%
\typeout{** loaded for the language `#1'. Using the pattern for}%
\typeout{** the default language instead.}%
\else
\language=\csname l@#1\endcsname
\fi
#2}}
\providecommand{\BIBdecl}{\relax}
\BIBdecl

\bibitem{arulkumaran2017brief}
K.~Arulkumaran, M.~P. Deisenroth, M.~Brundage, and A.~A. Bharath, ``A brief survey of deep reinforcement learning,'' \emph{arXiv preprint arXiv:1708.05866}, 2017.

\bibitem{mnih2015human}
V.~Mnih, K.~Kavukcuoglu, D.~Silver, A.~A. Rusu, J.~Veness, M.~G. Bellemare, A.~Graves, M.~Riedmiller, A.~K. Fidjeland, G.~Ostrovski \emph{et~al.}, ``Human-level control through deep reinforcement learning,'' \emph{nature}, vol. 518, no. 7540, pp. 529--533, 2015.

\bibitem{li2017deep}
Y.~Li, ``Deep reinforcement learning: An overview,'' \emph{arXiv preprint arXiv:1701.07274}, 2017.

\bibitem{lewis2012optimal}
F.~L. Lewis, D.~Vrabie, and V.~L. Syrmos, \emph{Optimal control}.\hskip 1em plus 0.5em minus 0.4em\relax John Wiley \& Sons, 2012.

\bibitem{sutton1999reinforcement}
R.~S. Sutton and A.~G. Barto, ``Reinforcement learning: An introduction,'' \emph{Robotica}, vol.~17, no.~2, pp. 229--235, 1999.

\bibitem{matignon2006reward}
L.~Matignon, G.~J. Laurent, and N.~Le~Fort-Piat, ``Reward function and initial values: Better choices for accelerated goal-directed reinforcement learning,'' in \emph{International Conference on Artificial Neural Networks}.\hskip 1em plus 0.5em minus 0.4em\relax Springer, 2006, pp. 840--849.

\bibitem{engel2014line}
J.-M. Engel and R.~Babu{\v{s}}ka, ``On-line reinforcement learning for nonlinear motion control: Quadratic and non-quadratic reward functions,'' \emph{IFAC Proceedings Volumes}, vol.~47, no.~3, pp. 7043--7048, 2014.

\bibitem{weber2023steady}
D.~Weber, M.~Schenke, and O.~Wallscheid, ``Steady-state error compensation for reinforcement learning-based control of power electronic systems,'' \emph{IEEE Access}, 2023.

\bibitem{franccois2018introduction}
V.~Fran{\c{c}}ois-Lavet, P.~Henderson, R.~Islam, M.~G. Bellemare, J.~Pineau \emph{et~al.}, ``An introduction to deep reinforcement learning,'' \emph{Foundations and Trends{\textregistered} in Machine Learning}, vol.~11, no. 3-4, pp. 219--354, 2018.

\bibitem{lilicrap2016continuous}
T.~Lilicrap, J.~Hunt, A.~Pritzel, N.~Hess, T.~Erez, D.~Silver, Y.~Tassa, and D.~Wiestra, ``Continuous control with deep reinforcement learning,'' in \emph{International Conference on Learning Representation (ICLR)}, 2016.

\bibitem{elfwing2018sigmoid}
S.~Elfwing, E.~Uchibe, and K.~Doya, ``Sigmoid-weighted linear units for neural network function approximation in reinforcement learning,'' \emph{Neural networks}, vol. 107, pp. 3--11, 2018.

\bibitem{lin2020comparison}
Y.~Lin, J.~McPhee, and N.~L. Azad, ``Comparison of deep reinforcement learning and model predictive control for adaptive cruise control,'' \emph{IEEE Transactions on Intelligent Vehicles}, vol.~6, no.~2, pp. 221--231, 2020.

\bibitem{ge2021numerically}
Q.~Ge, Q.~Sun, S.~E. Li, S.~Zheng, W.~Wu, and X.~Chen, ``Numerically stable dynamic bicycle model for discrete-time control,'' in \emph{2021 IEEE Intelligent Vehicles Symposium Workshops (IV Workshops)}.\hskip 1em plus 0.5em minus 0.4em\relax IEEE, 2021, pp. 128--134.

\bibitem{zheng2024highway}
Z.~Zheng, X.~Liu, and Y.~Lin, ``Highway discretionary lane-change decision and control using model predictive control,'' \emph{arXiv preprint arXiv:2402.17524}, 2024.

\end{thebibliography}

\end{document}